# How to Build an RSS Feed using ASP

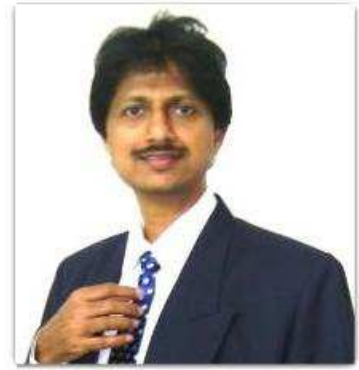

**By- Umakant Mishra, Bangalore, India**

umakant@trizsite.tk, http://umakant.trizsite.tk

**Contents**

## 1. Introduction to RSS

RSS or Really Simple Syndication is an open standard for distribution of web contents. The websites use RSS to distribute their web content and the web clients use RSS to fetch the recent updates from different websites. RSS has become a popular standard used by millions of websites and web users for sharing of web content. The content that is popularly disseminated through RSS are blogs, articles, news, events, discussions etc.

RSS is a XML based format. The Current popular version of RSS is RSS version 2.0. The previous version RSS 1.0 was known as Rich Site Summary. The versions are all backward compatible. I will not write more on the introduction as I had already done it in my previous article "The mechanics of Implementing RSS" (http://papers.ssrn.com/abstract=1974987). I will rather discuss on the practical aspects of how to build an RSS feeder and use the classical ASP (Active Server Pages) to code a RSS feed.

## 2. What is the benefit of creating RSS

Creating RSS feeds benefits both publishers and readers. As the objective of the publishers is to make their content available easily to its readers or audience and the objective of the readers is to get/find the content/updates easily, the RSS is targeted to achieve both these benefits. The websites display its latest updates through RSS feeds, which are used by other sites, search engines or users to access the latest content easily.

Using RSS is of tremendous benefit to the search engines, crawlers and bots that keep track of all the additions in the web. If there was no RSS then a search engine would have to do enormous processing to find what are the latest changes in a website. The search engine would have to crawl each of the pages and compare its hash with its previous versions in order to find the changes in a website. This mechanism is poor and ineffective for larger sites as the search engine would have to process thousands of comparisons. RSS has simplified the issue as the search engine gets all the updates of the site just from one single place, sorted and organized, without searching anywhere in the site.

⇨ Publishers (websites) are benefited as the feeds permit instant distribution of the updates in their content.
⇨ Search engines easily get the updates without crawling through the whole site
⇨ Magazines and similar sites easily get the updates on specific topics
⇨ Implementing RSS feeds can significantly improve the traffic to a website.
⇨ Advertizers also benefit as RSS feeds don't suffer from the drawbacks traditional marketing channels, such as, spam filters, delayed distribution, search engine ranking and general inbox noise.
⇨ Consumers are benefited as their subscription to feeds makes it possible to review a large amount of content in a very short time.
⇨ Web users are benefited in getting latest updates on the topic of their interest.

# 3. How to Code the RSS Feed

The purpose of adding an RSS feed to your site is to show if anything new is added to the site. For example, if a new article or blog or news item is added to your site that should automatically appear in the RSS feed so that the visitors/ RSS readers will automatically get updated about this new addition. The RSS feed is also called RSS channel.

As I mentioned above, the RSS feed is an XML file describing the latest updates of the site. The XML file should display the content according to RSS specifications. In this article we will go by the Current version of RSS, i.e., RSS 2.0 which is in existence since more than a decade.

There are two main elements of the RSS XML file, one is the header or channel element that describes the details about the site/feeder and other is the body or item element that describes the consists of individual articles/entries updated in the site. As the format of the RSS feed file is pretty simple, it can be coded in any language, ASP, PHP and anything of that sort.

# 4. Coding RSS Channel Element in ASP

The first part of an RSS feed consists of channel element or the information about the website or the news distributor. The most important information presented here are title, link and description. Other information may be added optionally such as language, copyright, webmaster, pubdate, lastBuildDate, category, ttl, image etc.

As the RSS is coded in XML format the first thing is to declare the XML version and the encoding whether utf8 or ISO. So the first line should appear like below.

> `<?xml version="1.0" encoding="utf-8"?>`
> or,
> `<?xml version="1.0" encoding="iso-8859-1"?>`

The next thing is to declare the RSS version and name spaces. As we are following RSS 2.0 we can declare it in short like this,

> `<rss version="2.0">`

Or we can add other namespaces as below.

> `<rss version="2.0"`
> `  xmlns:dc="http://purl.org/dc/elements/1.1/"`

```
xmlns:content="http://purl.org/rss/1.0/modules/content/"
xmlns:admin="http://webns.net/mvcb/"
xmlns:rdf="http://www.w3.org/1999/02/22-rdf-syntax-ns#">
```

The next is to add channel information or the source website information which should include a title, a link and some description at the minimum. The charm of RSS 2.0 specification is that it allows additional information in the channel element by adding new tags as shown above. So you can optionally add fields like the administrator's email, the error reporting email and other information that you may feel necessary for your RSS feed. The following is the XML code that I used to create the RSS feed for my website.

```
<channel>
    <title>the JournalSite</title>
    <link>http://www.journalsite.tk</link>
    <description>
    Instant eJournal for Self Publishing Authors!
    </description>
    <language>en-us</language>
<creator>Umakant Mishra (umakant@journalsite.tk) </creator>
<copyright>Copyright 2011-2013 JournalSite, All Rights Reserved.</copyright>
    <image>
    <url>http://journalsite.tk/journalsite/library/images/logo-3.png</url>
    <title>the JournalSite</title>
    <link>http://www.journalsite.tk</link>
    </image>
    <docs>http://www.rssboard.org/rss-specification</docs>
```

One more data to be added is lastbuilddate (the date and time of RSS creation) and a TTL (Time to Live). This requires some ASP code to format the date and time to present in accordance with the RSS specification.

```
<%
CurrentDate = Now()                          'Current time
'add a 0 before single digits to make it double digit figure
CurrentHour = Hour(CurrentDate)
if CurrentHour < 10 then CurrentHour = "0" & CurrentHour
CurrentMin = Minute(CurrentDate)
if CurrentMin < 10 then CurrentMin = "0" & CurrentMin
CurrentSec = Second(CurrentDate)
if CurrentSec < 10 then CurrentSec = "0" & CurrentSec
```

```
CurrentDateTime = WeekdayName(Weekday(CurrentDate), TRUE) & ", "
& Day(CurrentDate) & " " & MonthName(Month(CurrentDate), TRUE) & " "
& Year(CurrentDate) & " " & CurrentHour & ":" & CurrentMin & ":" &
CurrentSec & " EST"
%>
<lastBuildDate><%=CurrentDateTime%></lastBuildDate>
<ttl>240</ttl>
```

The above is enough for any standard site. But as I mentioned above, additional information may be inserted into the channel section by using specific tags. Such as, administrator, managingEditor, webmaster, generator etc.

## 5. Coding RSS Item Element in ASP

After the channel information is presented the next job is to present the item information for each new item. Here one may have to decide how many entries you want to show in your RSS feed. In a typical site displaying 10 recent entries should be fine.

The body/item part mainly has two functions, one is to read the data from database and the other is to present data in XML format. Assuming the data in your website is stored in a database, the reader part will read the data from the database. For the purpose we will use the commands to open and read the database. As we are interested to fetch the most recent records it will be simple to sort the data in descending order before reading the database.

The code for opening the database and reading the database may vary depending on your data source, database name etc. In the following example I have stored data in Microsoft Access file named articles.mdb. The code opens the Microsoft Access file through ADODB connection and keeps the data ready for presentation.

```
<%     'this is the rss feed for articles only, comments are in a separate rss
       set objcon=Server.CreateObject("ADODB.Connection")
       objcon.Open "Provider=Microsoft.Jet.OLEDB.4.0;Data Source=" & _
       server.MapPath("/database/articles.mdb")
       Set rsArticles=Server.Createobject("ADODB.Recordset")
       rsArticles.Open "select * from articles where final=true and disabled=false
order by lastupdate desc",objcon,adOpenKeyset,adLockOptimistic
%>
```

In the above code the data is read from the database "article.mdb" and made available through a recordset named rsArticles. It may be understood that the Articles file contains detailed information about each article.

The next part of the RSS feed is to display the actual RSS feeds for each individual item or article. The important fields here are title, link and description. The other informations may be added optionally. In the example below I have added author name and publication date.

```
<%      i=0
        while not rsArticles.eof  and i<10 'display recent 10 records
        i=i+1
        mDescription = rsArticles("description")   'description
        mArticleName = rsArticles("articlename")        'title of the article
        mLastUpdate = rsArticles("lastupdate")   'date updated
        mHour = Hour(mLastUpdate)                       'formatting date
        if mHour < 10 then mHour = "0" & mHour
        mMinute = Minute(mLastUpdate)
        if mMinute < 10 then mMinute = "0" & mMinute
        mSeconds = Second(mLastUpdate)
        if mSeconds < 10 then mSeconds = "0" & mSeconds
mLastUpdateTime = WeekdayName(Weekday(mLastUpdate), TRUE) & ", " &
Day(mLastUpdate) & " " & MonthName(Month(mLastUpdate), TRUE) & " " &
Year(mLastUpdate) & " " & mHour & ":" & mMinute & ":" & mSeconds & " +0530"
        %>
        <item>
        <title>
'<![CDATA['<%=rsArticles("authorname")%>']]>' published an article
'<![CDATA['<%=mArticleName%>']]>'
        </title>
        <link>
http://journalsite.tk/journalsite/articles/viewarticle.asp?articleid=<%=rsArticles("art
icleid")%>
        </link>
        <description>
        ABSTRACT: '<![CDATA['<%=mDescription%>']]>'
        </description>
        <pubDate><%=mLastUpdateT%></pubDate>
        </item>
```

```
<%      rsArticles.MoveNext
        wend
        rsArticles.close
        set rsArticles=nothing
        set objcon=nothing
        %>
        </channel>
</rss>
```

The above code reads 10 records from the file (i<11) and displays the title, link, description and pubdate for each article. Mark here the title in RSS is not same as the title of the article. The title in the RSS is like "authorname published an article articlename" which is composed in the above code. The fields generally used in item section of an RSS are as follows.

# title- title of the article, news or item

# link- URL of the article, news or item

# author – name or email of the author

# comments – if any

# enclosure- if any media object is attached to the item then specify the URL of media object

# guid- a string that uniquely identifies the item, can be the unique URL of the item

# pubdate- the date and time of publication

# 6. Issues involved in XML conversion

## Issues on date specification

One problem with the date is that usually the articles fed online are saved with the server date and time. The server might have been configured a local time where the server is located or might have been configured with some other time. So one has to see the server time and specify the publication time accordingly.

The date format should be either of the following types. Any other date format may lead to error or may not be recognized by the RSS reader. The date should be a two-digit figure. So if the date is less than 10 then a 0 (zero) is to be added as suffix, 01, 02, 03 etc.

Sat, 30 Jun 2013 15:21:36 GMT
Sat, 30 Jun 2013 15:21:36 EST
Sat, 30 Jun 2013 15:21:36 +0530

The last of the above three formats specifies the time as ahead or behind the GMT. In case of India it will be +0530 which means 5 hours 30 minutes ahead of GMT.

It is important to check that the time after calculation to ensure that the calculation is not done wrongly to show a future time.

**Issues on Email specifications**

The email should be a valid email address as per RFC 2822. For example it can be like:

umakant@trizsite.tk
umakant@trizsite.tk (Umakant Mishra)

The email may be entered with dc:creator element in order to create the authorship without revealing the email address.

**Invalid characters in the feed**

Some characters like -, &, >, < etc. are not allowed inside the XML. There are a couple of options. One option is to replace these characters with their equivalent codes before they are inserted into the RSS feed. For example,

<title>A < B</title> will display A<B

If the data is extracted from a database it may be replaced programmatically by using ASP, PHP or other programming language. For example if there are &,<,> symbols in the description field you can replace it with equivalent code in ASP as below.

mDescription = Replace(mDescription, "&", "&")
mDescription = Replace(mDescription, "<", "<")
mDescription = Replace(mDescription, "&", ">")

However, if you are including descriptions (or varchar data types) there may be many invalid type of characters and replacing each of them may be tedious or problematic. The other alternative is to accept whatever data is there as it is by using a cData tag as below. The cData tag starts with '<![CDATA[' and ends with ']]>' and anything in between these two tags are displayed as it is.

'<![CDATA['<%=mDescription%>']]>'

However you cannot use both the above methods, i.e., using Hex code and using cData. In that case the output will be misleading.

If there are curly quotes and curly apostrophes then change the xml encoding from "utf-8" to the following to solve the problem.
<?xml version="1.0" encoding="iso-8859-1"?>

**Language Specifications**

The language should be in compliance to the W3C format. It cannot be the spoken name of the language like "English". It has to be the language code such as, "en-us" and the like. Here the language code "en" is attached with a country code "us" by a hyphen (-).

# 7. Viewing the Feed Output

The RSS feed may be viewed by a RSS feed reader or RSS aggregator. There are many specialized feed reader programs which you can customize to display the content of your interest based on keywords. You can also customize whether to display the full content or only the headlines. Even many content management systems (CMS) provide RSS reading features to collect updates from different sites as a cron job.

Many browsers like Internet Explorer, Opera, Firefox etc. automatically detect RSS feeds and support formatted browsing of RSS feeds. The following screenshot displays Opera's style of displaying RSS feeds for the above code.

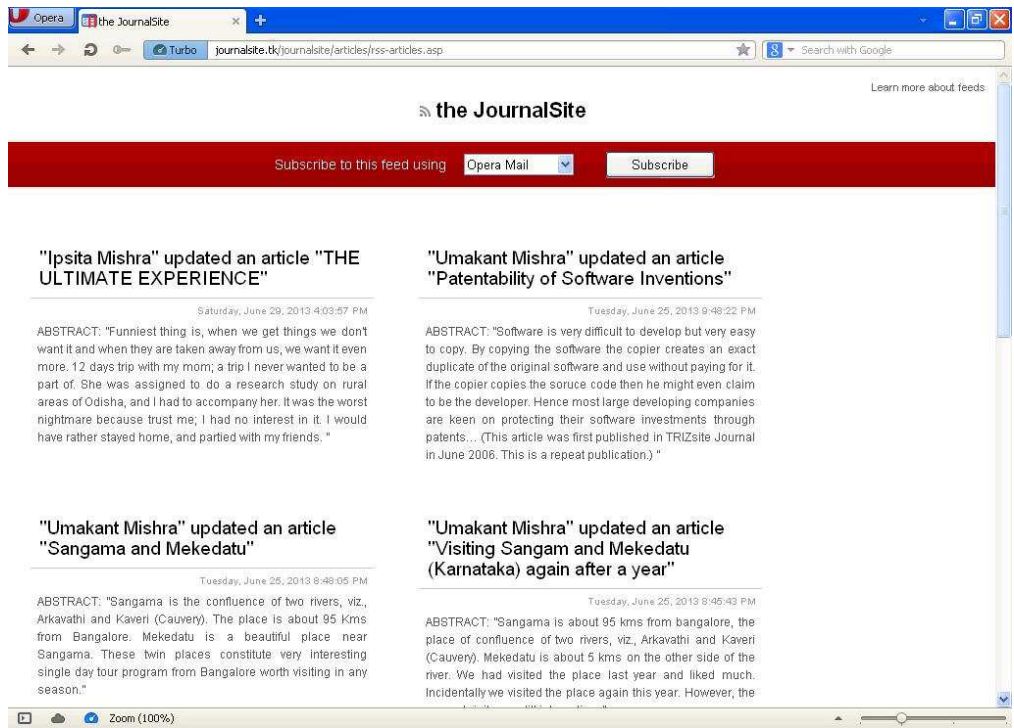

There are many RSS viewing software. Some of the free software are FeedDemon, SharpReader, SurfPack, FeedReader, RSSReader etc.

SurfPack ([www.surfpack.com](www.surfpack.com))- is a free desktop application that gathers up information from favorite feeders to present in newsfeeds or HTML or DHTML pages. Surfpack can feature search tools, Current weather conditions, LiveJournal diaries, humor and other dynamically updated modules through RSS, Atom and XML formats.

RSSReader ([www.rssreader.com)-](www.rssreader.com) is a Freeware that collects news from various RSS and Atom feeds in the background at user configurable intervals. Runs on all versions of Windows.

FeedDemon – ([www.feeddemon.com](www.feeddemon.com)) is a free Windows based software for Reading RSS feeds from hundreds of websites. It has features for tagging keywords to items for classification, searching keywords and download the target pages automatically etc.

SharpReader ([www.sharpreader.net](www.sharpreader.net)) is an RSS/Atom aggregator for Windows based on .NET framework. In most cases the ShapReader will auto-discover the RSS for a site. Alternatively you can type the RSS URL into the SharpReader's address bar to open the.

Google reader was a popular reader which will not be supported anymore from Google. Other popular readers are NewsGator, MyYahoo, Bloglines, Pageflakes, Netvibes etc. You can find a longer list of RSS readers in [http://www.rss-specifications.com/rss-readers.htm](http://www.rss-specifications.com/rss-readers.htm).  All these readers work very similar to email programs. They will read the unread entries from the RSS channels and fetch them for the client.

## 8. Validating the Code

There are many reasons for getting errors while parsing the code by a RSS reader. There may be errors in XML formatting or incompatibility with RSS specification. Hence it is always better to validate the code by a validating engine. Using feedvalidator is one of the best options that is available free. You can download feedvalidator or use feedvalidator.org to test the feed online.

The site feedvalidator.org is extremely powerful and helpful in code validation and error detection. Feedvalidator will validate your code online for free and give an excellent report on mistakes and possible solutions. It will display the line numbers that contain errors and highlight those lines for easy detection. It contains a large set of documents (http://feedvalidator.org/docs/) to display on specific errors.

After seeing the errors in your code it is the time to modify your code to remove errors. If the error is coming because of invalid data then you may consider editing data (if occasional) or adding a filter (in the data entry screen). However, if it is not easy or meaningful to edit data then best option is to use cData section (as in the above example) in order to accept any special and unsupported character as data and present as it is.